\documentstyle[aps,prl,epsf,multicol]{revtex}
\draft
\newcommand{\be}{\begin{equation}}
\newcommand{\ee}{\end{equation}}
\newcommand{\bd}{\begin{displaymath}}
\newcommand{\ed}{\end{displaymath}}
\newcommand{\ba}{\begin{eqnarray}}
\newcommand{\ea}{\end{eqnarray}}

\def \theequation{\arabic{equation}}

\def\av#1{{\left\langle#1\right\rangle}}
\def\gsim{\lower3pt\hbox{$\stackrel{>}{\scriptstyle \sim}$}} %greater, sim.
\def\lsim{\lower3pt\hbox{$\stackrel{<}{\scriptstyle \sim}$}} %less, sim.
\def\ave{\overline{\varepsilon}}

\begin{document}

\def\runtitle{Local properties of extended self-similarity in 3D    
turbulence}
\def\runauthor{Daigen Fukayama, Tohru Nakano, A. Bershadskii and T. Gotoh }
\title
{
Local properties of extended self-similarity in 3D turbulence}

\author
{
Daigen Fukayama,$^{1}$
Tohru Nakano,$^{1}$
Alexander Bershadskii$^{1,2}$ and Toshiyuki Gotoh$^{3}$
}

\address
{
$^1$ Department of Physics, Chuo University, Tokyo 112-8551, Japan,\\
$^2$ Machanaim Center, P.O. Box 31155, Jerusalem 91000, Israel,\\
$^3$ Department of Systems Engineering, Nagoya Institute of Technology, 
Nagoya 466, Japan.\\
}

\date{Version 3.0, \today}

\maketitle

\begin{abstract}\noindent
Using a generalization of extended self-similarity we have studied local
scaling
properties of incompressible homogeneous isotropic 3D turbulence in a direct
numerical simulation.  We have found that these properties are
consistent with lognormal-like behavior of the velocity increments
with moderate amplitudes for space scales $r$ beginning from Kolmogorov
length
$\eta$ up to
the largest scales, and in the whole range of the Reynolds numbers: $50 \leq
R_{\lambda}
\leq 459$.  The locally determined intermittency exponent $\mu(r)$ varies
with $r$;
it has a maximum at scale $r=14 \eta$, independent of $R_{\lambda}$.
\pacs{PACS numbers: 47.27.Ak, 47.27.Jv, 47.27.Gs, 05.20.Jj}
%\noindent PACS numbers: 47.27.Ak, 47.27.Jv, 47.27.Gs, 05.20.Jj

\end{abstract}

\begin{multicols}{2}
So-called extended self-similarity (ESS) in incompressible turbulent flows is
intensively studied in
recent years (see, for instance,~\cite{benzi1,benzi2,dh,sreen,cao,gr,ber,fu}
and references therein).  The ESS implies
scaling relation between moments of different order.  For example, for
absolute value of longitudinal velocity increments over separation $r$ in the
inertial range the ESS means
\be
  \av{u_r^q} \sim \av{u_r^p}^{\rho(q)/\rho(p)},  \label{0}
\ee
where the scaling exponent $\rho(q)$ is some function of $q$.  It
is shown
in numerous experiments and numerical simulations that the range of
applicability of
the ESS is substantially larger than that for ordinary self-similarity, and
the ESS
can exist even for situations where the ordinary self-similarity cannot be
observed at
all.

The aim of the present Letter is twofold.  The first is to study local
properties of
ESS defined as
\be
     \av{ u_r^p} \sim  \av{u_r^3}^{\zeta_p(r)},  \label{1}
\ee
where $\zeta_p(r)$ are depending on $r$.  The local approach allows to
develop some
old ideas.  At past a representation $\zeta_p=p/3-\mu p(p-3)/18$ derived from a
lognormal model, was shown to hold for the values averaged over an inertial
interval
of certain extension.   Here we will show that the representation holds locally
(i.e. with
$\zeta_p(r)$ and $\mu(r)$ depending on $r$) in the above defined sense,
ranging from
the dissipative Kolmogorov length $\eta$ to integral scale for any observed
Reynolds
number in direct numerical simulations up to $R_{\lambda}=459$.

The second aim is to examine the implication of the variation of the local
intermittency exponent $\mu(r)$.  It is often mentioned that there
is no characteristic length in turbulence, so that the structure functions
obey a
power law in the inertial region and the associated scaling exponents are
independent of scale.  However, a recent study of turbulence reveals the
existence of the structures~\cite{j}.  A question, then, naturally arises as to
whether there may be any characteristic length ascribed to the
structures~\cite{km}.
The peculiar variation of $\mu(r)$ with respect to $r$, observed in the Letter,
indicates that there is a certain length affecting the ESS.

The She-Leveque model~\cite{sl} is also very popular in the last years.  For
this
model the local exponent is expressed in a general form:
\be
    \zeta_p(r)=\frac{p}{3}(1-\gamma)+\frac{\gamma}{1-\beta}
    \left[ 1-\beta^{p/3}\right],  \label{2}
\ee
where $\beta$ and $\gamma$ can be also supposed to depend on $r$; $\beta(r)$
and
$\gamma(r)$ could be in principle evaluated by plotting the local exponent
$\zeta_p(r)$ against $p$ with $r$ fixed.  Since here we focus the lower order
structure functions which are computed more reliably than the higher order
ones, the comparison of the data with the formula is done using a simpler
lognormal model with one parameter.  If one could compute the higher order
structure functions, the comparison should be made with the She-Leveque model.

Let us begin with the derivation of useful formulae on the basis of the
lognormal distribution of the dissipation rate $\varepsilon_r$ averaged over
spheres
of radius $r$:
\be
   P(\varepsilon_r)=\frac{\varepsilon_r^{-1}}{\sqrt{2\pi \sigma^2}}
      \exp \left( -\frac{(\ln \varepsilon_r-a)^2}{2 \sigma^2} \right),
\label{3}
\ee
from which we obtain
\be
   \frac{\av{\varepsilon_r^q}}{\av{\varepsilon_r}^q}
    ={e}^{\sigma^2 q(q-1)/2} \label{4}
\ee
that results in a parameter-independent type of ESS~\cite{benzi1,benzi2} of
turbulent energy dissipation
\be
   \frac{\av{\varepsilon_r^q}}{\av{\varepsilon_r}^q}
   =\left( \frac{\av{\varepsilon_r^p}}{\av{\varepsilon_r}^p} \right)
      ^{\frac{q(q-1)}{p(p-1)}}. \label{5}
\ee
What is the equivalent relation for the velocity increment?  
According to the refined similarity
method~\cite{o,k,my,p,tv} $u_r \sim (r \varepsilon_r)^{1/3}$ in the inertial
region, while $u_r \sim r \varepsilon_r^{1/2}$ in the dissipative region.  
If we take regions of scale $r$ between the inertial and dissipative
regions, some are described by the former relation, while the others are by the
latter one, so that $u_r$ is related to $\varepsilon_r$ in a probabilistic way.
For the sake of simplicity we assume a general relation
\be
   u_r= c(r) \varepsilon_r^{1/\alpha(r)} \label{6}
\ee
in a mean sense.  Here $\alpha(r)$ is a function of $r$ and
coefficient $c(r)$
does not necessarily scale with $r$.  Substituting (\ref{6}) into (\ref{5})
yields
\be
   \frac{\av{u_r^q}}{\av{u_r^{\alpha}}^{q/\alpha}}
   =\left( \frac{\av{u_r^p}}{\av{u_r^{\alpha}}^{p/\alpha}} \right)
      ^{\frac{q(q-\alpha)}{p(p-\alpha)}}.  \label{7}
\ee
Note that this expresses how the structure functions of different order are
related to each other with value of $r$ fixed.
This relation can be considered as a functional equation and a solution to this
equation is
\be
   {\av{u_r^q}} \sim \av{u_r^{\alpha(r)}} ^{q/\alpha(r)+b(r) q(q-\alpha(r))},
     \label{8}
\ee
where $b(r)$ is an arbitrary function of $r$.  Relation (\ref{8}) is a
generalization of ordinary ESS~\cite{benzi1,benzi2}.  The difference between
ordinary ESS and relation (\ref{8}) is that parameters used in (\ref{8}) can
depend
on $r$.  Therefore we will call this type of ESS as extended {\it local}
self-similarity (ELSS).

To be consistent with the present data processing we express the $q$th order
structure function in terms of 3rd order structure function as (\ref{1}).
Making use of (\ref{8}), we are led to
\be
   \zeta_q(r)=\frac{q}{3}-\frac{\mu(r)}{18} q(q-3),  \label{10}
\ee
where
\be
   \mu(r)=-\frac{6 b(r) \alpha(r)}{1+\alpha(r)(3-\alpha(r))b(r)}.  \label{11}
\ee
In the frame of ELSS the exponent $\zeta_q$ depends also on $r$, and below we
compare (\ref{1}) and (\ref{10}) with data of DNS for
different values of $r$.  It should be noted that the ELSS expression
(\ref{10})
holds for any value of $\alpha(r)$, so that the expression can be compared
with the data for any scale separation without paying attention to which region is
being considered.  

We have performed a series of direct numerical simulations (DNS's) 
of incompressible homogeneous isotropic turbulence using a resolution 
up to $1024^3$. Reynolds numbers range from 50 to 459~\cite{gf}.
The random force is statistically homogeneous, isotropic and Gaussian white, 
and applied to the band $ 1 \lsim k \lsim 3$ in which the forcing spectrum 
is constant.  The code uses the pseudo spectral method and the 4th 
order Runge-Kutta-Gill one. Initial conditions are  
Gaussian random velocity fields with the energy spectrum 
$E(k)\propto k^4\exp(-2(k/k_0)^2)$, and the resolution is 
$N=256^3$ for $R_{\lambda}=69$, $N=512^3$ for $R_{\lambda}=125,176,259$, 
and $N=1024^3$ for $R_{\lambda}=374, 459$. 
After about two eddy turnover times  
all the turbulent fields attained statistically steady states, 
which were confirmed by observing 
the time evolution of the total energy and enstrophy, and the skewness 
of the longitudinal velocity derivative.
For $R_{\lambda}=459$ run, the Reynolds number was gradually increased 
through two steady states. 
The condition $k_{\max}\eta >1$ for the resolution of DNS is satisfied
for most runs, but that of $R_{\lambda}=459$ is slightly less than unity. 
The statistical averages were taken as the time average over tens of
turnover times for lower Reynolds numbers and over a few turnover times 
for the higher Reynolds numbers; over 45 samples during 2.9 eddy
turnover times for $R_{\lambda}=374$ and over 31 samples during 1.4
eddy turnover times $R_{\lambda}=459$.  Computations with $R_{\lambda}
\leq 259$ have been done using a vector  parallel machine with 16
processors, Fujitsu VPP700E at RIKEN, and those  for higher
$R_{\lambda}$, using Fujitsu VPP5000/56 with 32 processors  at Nagoya
University Computation Center.

Now turn to the data analysis.  Figure 1 is a plot of $\av{u_r^2}/(\ave
\eta)^{2/3}$
against $r/\eta$ for various values of Reynolds number, where $\ave$ is the
average
dissipation rate, and $\eta=(\nu^3/\ave)^{1/4}$ with molecular viscosity
$\nu$.
Here a straight solid line proportional to $r^{2/3}$ is inserted.  It is
remarkable
that all data points collapse on a single line in the dissipative region, which
indicates that all simulations are carried out with the good resolution at
small
scales.   Although the
slope of $\av{u_r^2}$ could be estimated for large Reynolds numbers as seen
from
Fig.1, the scaling exponents of higher order structure functions as well as low
order ones for small Reynolds numbers can be reliably evaluated only on the
basis of
the ESS method, i.e. by plotting $\av{u_r^p}$ against
$\av{u_r^3}$~\cite{benzi1,benzi2,dh}.

In order to know the $r$-dependence of $\zeta_p(r)$ for various Reynolds
numbers,
we prepare Fig.2, in which $\zeta_p(r)$ with $p=4, 6, 8$ are depicted for
$R_{\lambda}=69, 125, 259, 374$ and $459$.  (The eighth order structure function is
confirmed to converge statistically.)  Note that the data for scales larger
than
integral scales are not shown, because a universal property of turbulence is
not
expected in those data.  It is remarkable that there is a dip at about
$r/\eta
\sim 10$, and that it grows in depth with Reynolds number.  The exception is
the case
$R_{\lambda}=69$, where a dip does not appear.  As the scale increases
beyond the
dip, $\zeta_4(r)$ and
$\zeta_6(r)$ tend to approach constant values, although the corresponding
data for
$R_{\lambda}=259$ behave in a slightly different way from other cases.  For
$p=8$ the
situation is the same as for $p=6$, but the variation is
larger.  It is of interest to notice that for the largest Reynolds number $459$
the flat region is
observed in the interval
$100 \lsim r/\eta \lsim 300$, which may be identified with the inertial
region.  For
smaller Reynolds numbers it is a little difficult to find the flat region.  For
$R_{\lambda}=69$ we see the flat region in the interval $10 \lsim r/\eta
\lsim 30$,
and the corresponding slopes for $p \geq 4$ are larger than those for
$R_{\lambda}=459$. However, the flat region at smaller scales for
$R_{\lambda}=69$ is
different from one at larger scales for $R_{\lambda}=459$.

Before we compare the formula (10) with the data, the validity of
the lognormal distribution of the velocity increment should be ensured. 
The pdf of the intermediate amplitudes is certainly lognormal.  
In order to determine the range of the lognormality, 
we calculated a peak position $u_r^*(p)$ of $u_r^p P(u_r)$ at
$r/\eta=24$ for $R_{\lambda}=121$. 
The pdf is satisfactorily fitted by a lognormal curve for
$u_r^*(p=1.5) \leq u_r \leq u_r^*(p=6) $.  On the other hand, the pdf of
$\varepsilon_r$ is found to be lognormal in much wider interval.  If we employ the
same notation as above, the lognormality of
$\varepsilon_r$ holds at least in the interval of $\varepsilon_r^*(p=-4) \leq
\varepsilon_r \leq \varepsilon_r^*(p=6)$ for the same Reynolds number.  If the
refined similarity hypothesis $u_r^3 \sim r \varepsilon_r$ is assumed to
hold for any amplitude, the corresponding pdf of
$u_r$ should be lognormal for $u_r^*(p=-12) \leq u_r \leq u_r^*(p=18)$, which is
much wider than the observed lognormal interval.  The reason for the discrepancy
is that the refined similarity holds only for the intermediate amplitudes of
$\varepsilon_r$ in agreement with the observation\cite{p,tv,s1}.  Hence,
the use of the lognormal expression for the exponent (10) 
is completely justified for intermediate values of $p$.

In order to analyze a nature of the $r$-dependence of the local slope
$\zeta_p(r)$, we rewrite (\ref{10}) in the following form:
\be
   \frac{\zeta_p}{p}=\left( \frac{1}{3}+\frac{\mu(r)}{6} \right)
-\frac{\mu(r)}{18}p.
    \label{12}
\ee
What is the range of $p$?  To decide the range we calculated $\zeta_p(r)$ for
various values of $p$ at units of 0.1 at $r/\eta=24$ for $R_{\lambda}=121$, and
plotted $\zeta_p/p$ against $p$.  Although such a plot is not given here,
$\zeta_p/p$ is fitted by a straight line for $1 \leq p \leq 6$.   
The curve is deviated upward from the straight fitting line for $p \geq
7$ and downward for $0<p<1$.  Hence the comparison will be made in the range   
$1 \leq p \leq 6$,
but the intermittency coefficient $\mu(r)$ estimated there plays a
significant role to represent even the whole intermittency effect.

Figure 3a shows a curve $\zeta_p(r)/p$ vs. $p$ for several values of
$r/\eta$ with $R_{\lambda}=69$.  Straight lines in this figure are the best
fit lines
for $1 \leq p \leq 6$.  The data points for $p=7$, which are not included
for the
comparison, are slightly deviated from the lognormal lines as mentioned above. 
Intermittency index,
$\mu(r)$, can be calculated from this figure using the slope of the fitting
straight
lines or the intersection point of the fitting straight lines with vertical
axis
(cf.(\ref{12})).  The calculated values of $\mu(r)$ are shown in the inset
of the
Fig.3a ($\mu_1$ corresponds to calculations using the 'slope' method, whereas
$\mu_2$ are given by the 'intersection' method.)  In the plot of $\mu(r)$ 
Taylor microscale $\lambda$ and an integral scale $L$ are marked for
convenience.
(We have
confirmed the well-known prediction~\cite{tl}
$\lambda/\eta=15^{1/4} R_{\lambda}^{1/2}$ and
$L/\eta \sim R_{\lambda}^{3/2}$.)
For Reynolds numbers $259$ and $459$ we have obtained similar pictures (
Figs.3b and 3c).   Other Reynolds numbers give the same result.

As seen from Figs.3a to 3c, $\mu(r)$ substantially depends on $r$ with a
typical ^^ ^^ two-maxima" shape.  Even for $R_{\lambda}=69$ in Fig.3a there
is a
small peak around $r/\eta \sim 10$.  Local maximum of the $\mu(r)$ at smaller
scales exhibits a few interesting properties.  (I) For $R_{\lambda}=50$ the
maximum is
not observed, and the first appearance of the maximum occurs at $R_{\lambda}$
between 50 and 69.  (II) A position of this maximum
(normalized by $\eta$) is independent of Reynolds number,
and takes value $\ell_1 \approx 14 \eta$, which is actually the beginning of
the
dissipative region.  (III) Value of $\mu(\ell_1)$ scales with $R_{\lambda}$ as
$\mu(\ell_1) \sim R_{\lambda}^{0.26}$.
(IV) The scale $\lambda$ is located on the right descending hill.

The flat region where $\mu(r)$ is constant appears in between $r/\eta \sim
100$ and
$r/\eta \sim 300$ for the highest Reynolds number $R_{\lambda}=459$ (see
Fig.3c).
In such an interval $\mu(r) \approx 0.25$.  This value is consistent with
those known
in literature for observations corresponding to very large Reynolds
number~\cite{sk,po}.

It should be noted that strong tube-like vortices are believed to be
sparsely distributed in space in fully developed turbulence~\cite{j,mt}.  Those
vortices are frequently assumed as Burgers' vortex with mean radius $10
\eta$~\cite{mt}.  The energy dissipation takes place strongly around those
vortices.
Therefore extreme intermittency of the energy dissipation at this scale is
consistent
with the observation of the maximum of $\mu(r)$ at scale $r=14 \eta$.  On
the other
hand, the usual scaling region, where $\zeta_p(r)$ as well as $\mu(r)$ are
independent
of scale in a certain interval of scale, can be seen only for largest
Reynolds number,
$R_{\lambda}=459$ in the present work.  Such a scaling region is located
for $ r>\lambda$; in our simulation the scaling region starts from
approximately $2.5
\lambda$.

The work of T.G. presented here was supported by a Grant-in Aid for
Scientific Research (C-2) 12640118 from the Japan Society for the Promotion of
Science.  We are very grateful to RIKEN Computer Center and Nagoya University
Computer Center for their support.  T. Ochiai at NIT is also acknowledged
for his
assistance in numerical computation involved in this work.  One of the
authors A.B.
is grateful to C.H. Gibson and K.R. Sreenivasan for numerous discussions on the
problem.

\vspace*{-0.5cm}
%\begin{thebibliography}{99}

%\end{thebibliography}
%\newpage
\narrowtext
\begin{figure}
\epsfxsize=80mm
\epsfysize=53mm
\epsfbox{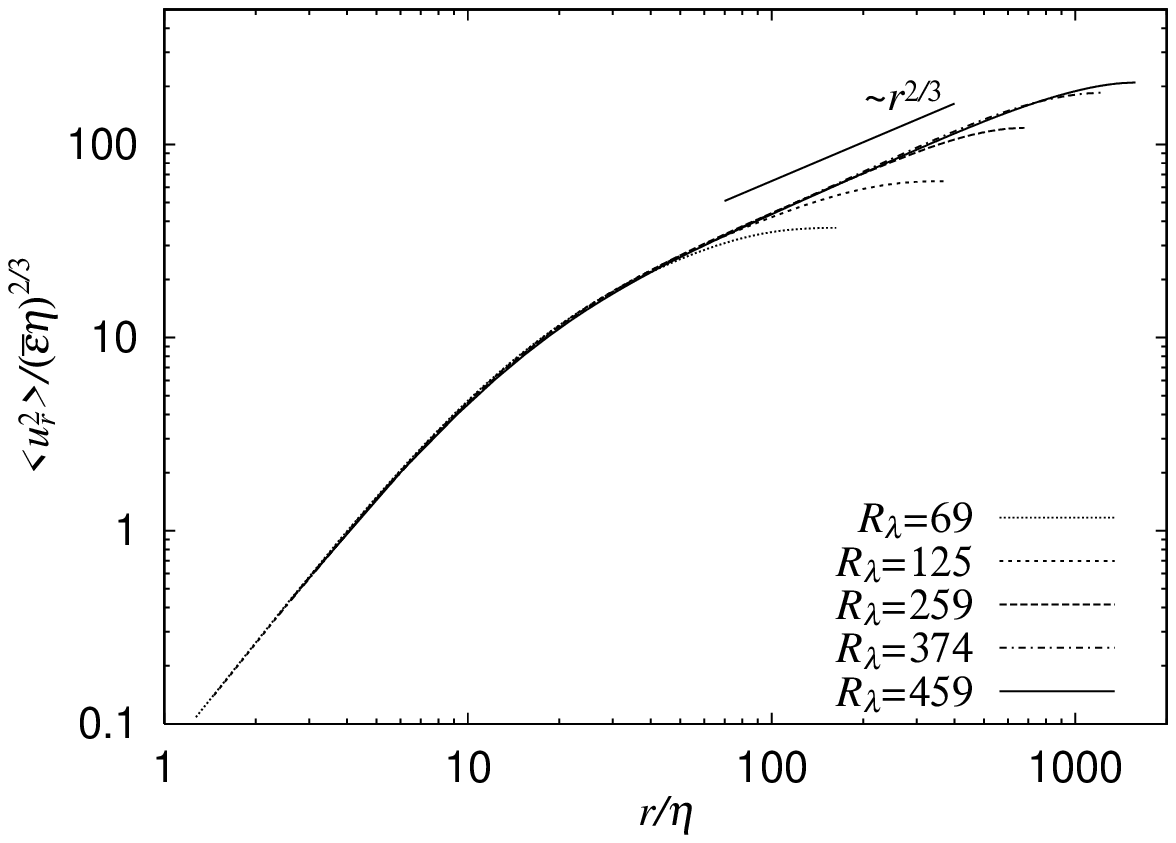}
\vspace{2mm}
\caption{A plot of $\av{u_r^2}$ divided by $(\ave \eta)^{2/3}$ vs.
$r/\eta$ for
various values of Reynolds number.  An inserted solid line is proportional to
$r^{2/3}$.  Notice that all data points collapse on a single curve in the
dissipative region.}
\label{fig:1}
\vspace{2mm}
\epsfxsize=80mm
\epsfysize=53mm
\epsfbox{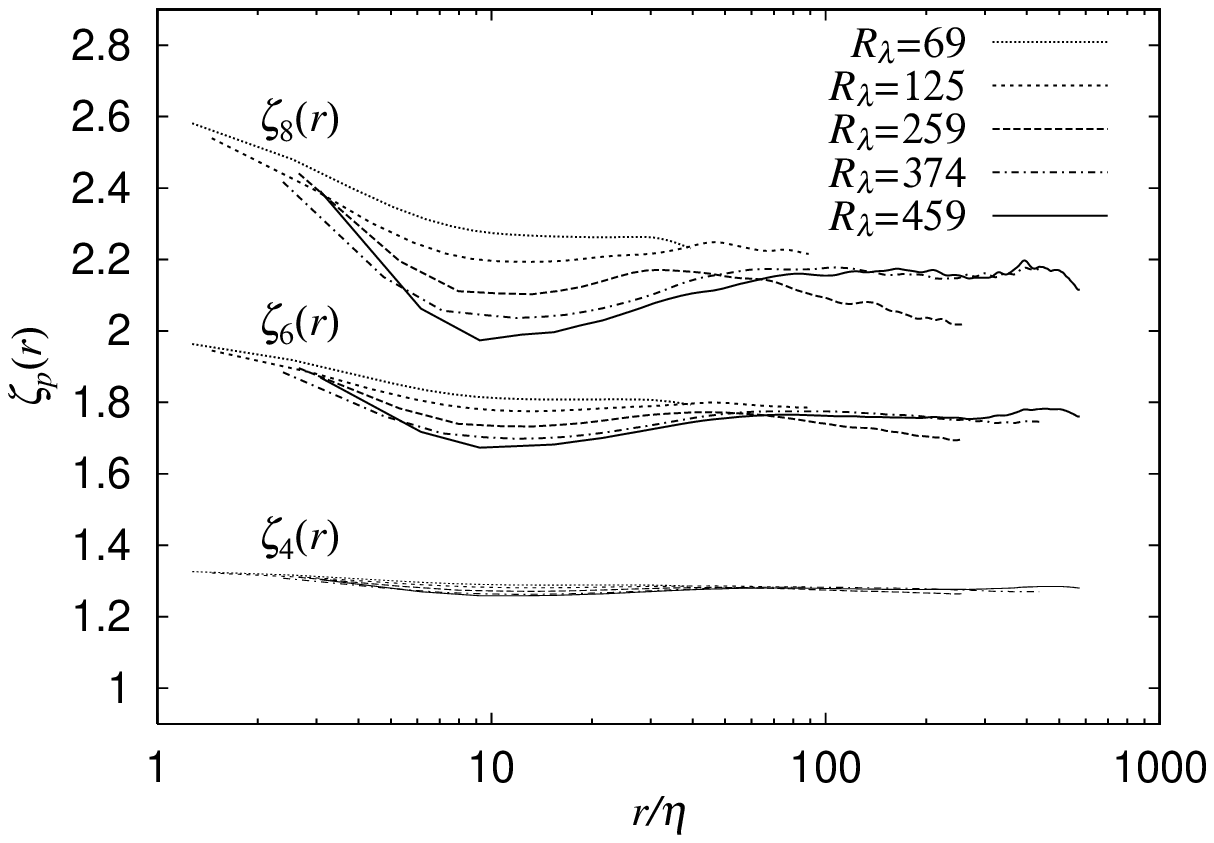}
\vspace{1mm}
\caption{
The ELSS exponents $\zeta_4(r), \zeta_6(r)$ and $\zeta_8(r)$ against
$r/\eta$ for various values of Reynolds number.  The data with scales larger
than
integral scales are deleted from the figure.
}
\label{fig:2}
\vspace{2mm}
\epsfxsize=80mm
\epsfysize=53mm
\epsfbox{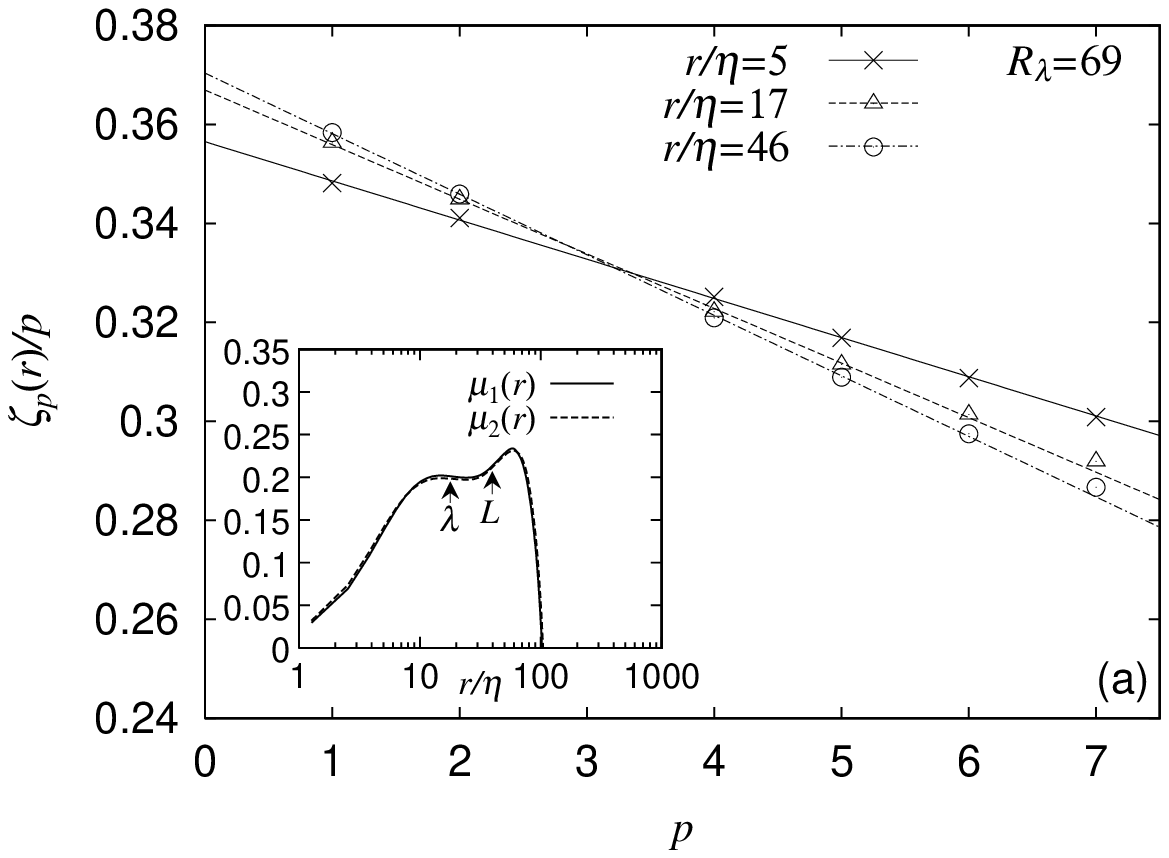}
\label{fig:3a}
\epsfxsize=80mm
\epsfysize=53mm
\vspace{2mm}
\epsfbox{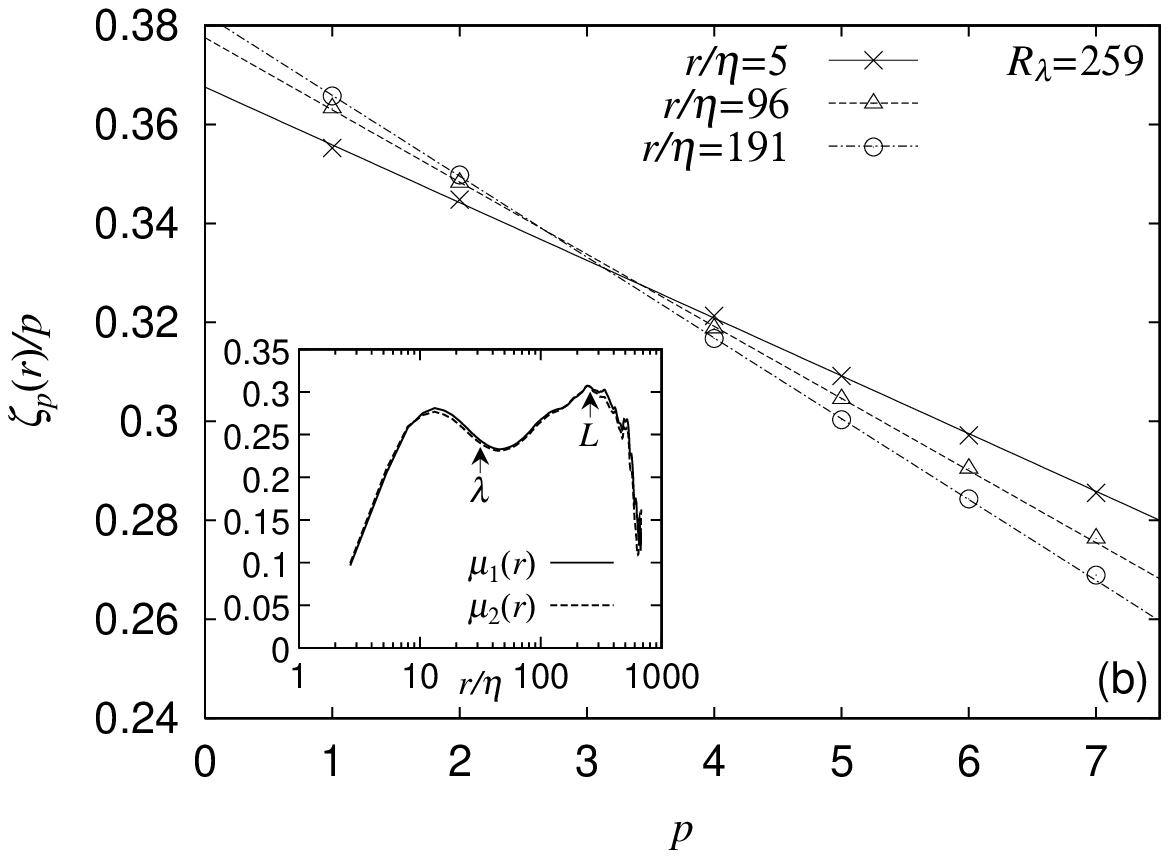}
\label{fig:3b}
\epsfxsize=80mm
\epsfysize=53mm
\vspace{2mm}
\epsfbox{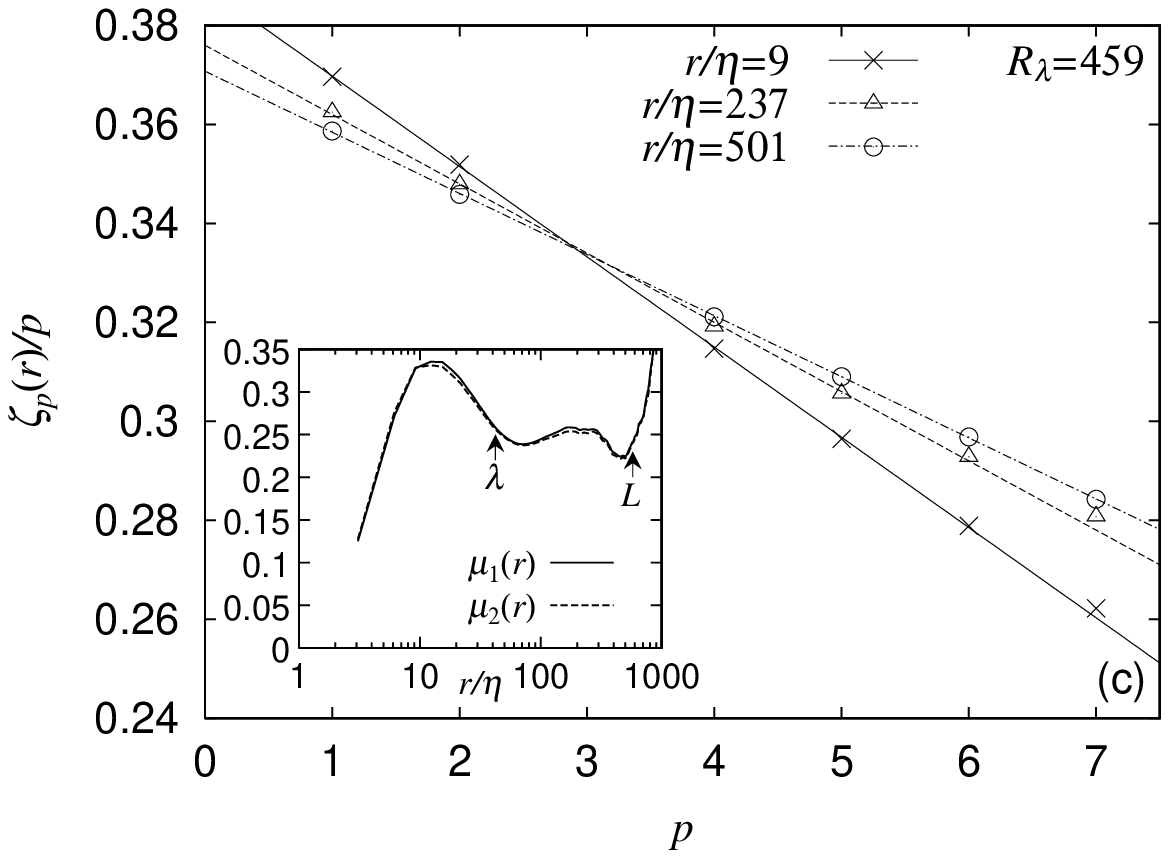}
\label{fig:3c}
\vspace{2mm}
\caption{
The ELSS exponents $\zeta_p(r)/p$ against $p$ obtained in the DNS
for
different values of $r$ and for different Reynolds numbers; (a)
$R_{\lambda}=69$, (b)
$R_{\lambda}=259$ and (c) $R_{\lambda}=459$.  Straight lines, which are the
best fit
line for $1 \leq p \leq 6$, indicate agreement of the data with the
representation
(12).  The inset shows local intermittency exponent $\mu(r)$ calculated
using the
data.  $\mu_1(r)$ corresponds to calculations using the 'slope' method
(described in
the text), and $\mu_2(r)$ does to those using the 'intersection' method.
Calculated Taylor length $\lambda$ and the integral length $L$ are marked for
convenience.
}

\end{figure}

\end{multicols}

\end{document}